\documentclass[aps,prl,a4,twocolumn,preprintnumbers,nofootinbib]{revtex4}
\newcommand{\bmat}{\left(\begin{array}}
\newcommand{\emat}{\end{array}\right)}
\newcommand{\be}{\begin{equation}}
\newcommand{\ee}{\end{equation}}
\newcommand{\bea}{\begin{eqnarray}}
\newcommand{\eea}{\end{eqnarray}}
\newcommand{\nn}{\nonumber}
\newcommand{\bi}{\bibitem}
\def\L{\left}
\def\R{\right}

\def\lsim{\raise0.3ex\hbox{$\;<$\kern-0.75em\raise-1.1ex\hbox{$\sim\;$}}}
\def\gsim{\raise0.3ex\hbox{$\;>$\kern-0.75em\raise-1.1ex\hbox{$\sim\;$}}}

\usepackage{graphicx,graphics}
\usepackage{dcolumn}
\usepackage{bm}
\begin{document}

\title{Dark Matter in $B-L$ Extended MSSM Models}
\author{S. Khalil$^{1,2}$}
\author{H. Okada$^{1}$ }
\affiliation{$^{1}$
Centre for Theoretical Physics, The British University in Egypt, El Sherouk City, Postal No, 11837, P.O. Box 43, Egypt.
}%
\affiliation{$^{2}$
Department of Mathematics, Ain Shams University, Faculty of Science, Cairo, 11566, Egypt.
}%

\begin{abstract}
We analyze the dark matter problem in the context of
supersymmetric $U(1)_{B-L}$ model. In this model, the lightest
neutalino can be the $B-L$ gaugino $\widetilde {Z}_{B-L}$ or the
extra Higgsinos $\widetilde{\chi}_{1,2}$ dominated. We compute the
thermal relic abundance of these particles and show that, unlike
the LSP in MSSM, they can account for the observed relic abundance
with no conflict with other phenomenological constraints. The
prospects for their direct detection, if they are part of our
galactic halo, are also discussed.
\end{abstract}

\maketitle

\section{Introduction}
Nonvanishing neutrino masses and the existence of nonbaryonic dark
matter (DM) are the most important evidences of new physics beyond
the Standard Model (SM). A simple extension of the SM, based on
the gauge group $G_{B-L} \equiv SU(3)_C \times SU(2)_L \times
U(1)_Y \times U(1)_{B-L}$, can account for current experimental
results of light neutrino masses and their large mixing
\cite{Khalil:2006yi}. In addition, the extra-gauge boson and
extra-Higgs predicted in this model have a rich phenomenology and
can be detected at the LHC \cite{s.khalil2}. It is worth
mentioning that several attempts have been proposed to extend the
gauge symmetry of the SM via one or more $U(1)$ gauge symmetries
beyond the hypercharge gauge symmetry
\cite{barger,Mohapatra:1980qe}.

Within supersymmetric context, it was emphasized that the three
relevant physics scales related to the supersymmetry, electroweak,
and baryon minus lepton $(B - L)$ breakings are linked together
and occur at the TeV scale \cite{s.khalil-a.masiero}. Indeed, it
was shown that radiative corrections may drive the squared mass of
extra $B-L$ Higgs from positive initial values at the GUT scale to
negative values at the TeV scale. In such a framework, the size of
the $B-L$ Higgs vacuum expectation value (VEV), responsible for
the $B-L$ breaking, is determined by the size of the right-haneded
Yukawa coupling and of the soft SUSY breaking terms.

In this paper, we consider the scenario where the extra $B-L$
neutralinos (three extra neutral fermions: $U(1)_{B-L}$ gaugino
$\widetilde {Z}_{B-L}$ and two extra Higgsinos
$\widetilde{\chi}_{1,2}$) can be cold DM candidates.  It turns out
that the experimental measurements for the anomalous magnetic
moment impose a lower bound of order $30$ GeV on the mass of
$U(1)_{B-L}$ gaugino $\widetilde {Z}_{B-L}$, while Higgsinos
$\widetilde{\chi}_{1,2}$ can be very light. We examine the thermal
relic abundance of these particles and discuss the prospects for
their direct detection if they form part of our galactic halo.

It is worth mentioning that assuming the lightest neutralino in
minimal supersymetric standard model (MSSM) as DM candidate
implies sever constraints on the parameter space of this model.
Indeed, in the case of universal soft-breaking terms, the MSSM is
almost ruled out by combining the collider, astrophysics and rare
decay constraints \cite{Ellis:2003cw} . Therefore, it is important
to explore very well motivated extensions of the MSSM, such as
SUSY $B-L$ model which provides new DM candidates that may account
for the relic density with no conflict with other phenomenological
constraints.

The paper is organized as follows. In section 2 we briefly review
the supersymmetric $U(1)_{B-L}$ model with a particular emphasis
on its extended neutralino sector. Section 3 is devoted for
computing the LSP annihilation cross section for $\widetilde
{Z}_{B-L}$, $\widetilde{\chi}_{1}$, and $\widetilde{\chi}_{2}$. In
section 4 we examine the possible constraints imposed by the
experimental limits of muon anomalous magnetic moment on the mass
of $\widetilde {Z}_{B-L}$. We discuss the relic abundance of these
DM candidates in section 5. We show that they can account for the
measured relic density without any conflict with other
phenomenological constraints. The direct detection rate of
$\widetilde {Z}_{B-L}$ and $\widetilde{\chi}_{1,2}$ is briefly
discussed in section 6. Finally we give our conclusions in section
7.

\section{$U(1)_{B-L}$ SUSY Model}
In $B-L$ extension of the MSSM, the particle content includes the
following fields in addition to the MSSM fields: three chiral
right-handed superfields ($N_i$), a vector superfield associated
to $U(1)_{B-L}$ ($Z_{B-L}$), and two chiral SM singlet Higgs
superfields ($\chi_1$, $\chi_2$).
This class of $B-L$ extension of the SM can be obtained from a
unified gauge theory, like $SO(10)$, with the following branching
rules for symmetry breaking: $SO(10)$ is broken down to Pati-Salam
gauge group: $SU(4)_c\times SU(2)_L\times SU(2)_R$ through the
vacuum expectation value (vev) of the Higgs: $(1,~1,~1)$ in $54_H$
or $210_H$ representation at GUT scale. Then Pati-Salam can be
directly broken down to $B-L$ model: $SU(3)_c\times SU(2)_L\times
U(1)_Y\times U(1)_{B-L}$ through the vev of the adjoint Higgs:
$(15,~1,~3)$ below GUT scale. Finally, the $B-L$ model is broken
down to $SU(3)_c\times SU(2)_L\times U(1)_Y$ at TeV scale as
mentioned above.
In this case, although the $U(1)_Y$ and $U(1)_{B-L}$ are exact
symmetries at high scale (larger than TeV), they are
non-orthogonal. This can be seen by noticing that the
orthogonality condition \cite{takeuchi} is not satisfied: $\sum_f
Y^f Y^f_{B-L}\neq 0$, where $Y^f$ and $Y^f_{B-L}$ are the
hypercharge and  the $B-L$ charge of the fermion particle $(f)$.
In this respect, there is a kinetic mixing between the gauge fields of $U(1)_Y$ and  $U(1)_{B-L}$. However, LEP results  \cite{m.carena} stringently constrain the corresponding mixing angle to be $\lsim$ $10^{-2}$.
Therefore, in our analysis, we neglect this small mixing and consider the following superpotential:
\begin{widetext}\bea {W} =
(Y_U)_{ij}Q_iH_uU^c_j+(Y_D)_{ij}Q_iH_dD^c_j+(Y_L)_{ij}L_iH_dE^c_j
+ (Y_{\nu})_{ij}L_iH_uN^c_j+(Y_N)_{ij}N^c_iN^c_j\chi_1%
+ \mu({H}_u{H}_d)+\mu'{\chi}_1{\chi}_2.
\label{super-potential-b-l}
\eea%
\end{widetext}
The $B-L$ charges of superfields appeared in the superpotential
$W$ are given in Table \ref{ub-l-charge}.
 \begin{widetext}
 \begin{table}[thb]
\begin{center}
\begin{tabular}{|c|c|c|c|c|c|c|c|c|c|c|} \hline
 & $l$ & $N$ & $E$ & $Q$ & $U$ & $D$ & $H_u$ & $H_d$ & $\chi_1$  & $\chi_2$
  \\ \hline
{ $SU(2)_L\times U(1)_Y$}
 & $ ({\bf 2}, -\frac{1}{2})$  &  $({\bf 1}, 0)$  &  $({\bf 1}, -1)$
 & $({\bf 2}, \frac{1}{6})$&  $({\bf 1}, \frac{2}{3})$
 &  $({\bf 1}, -\frac{1}{3})$& $({\bf 2}, \frac{1}{2})$& $({\bf 2}, -\frac{1}{2})$ &
$({\bf 1}, 0)$ &  $({\bf 1}, 0)$
  \\ \hline
 $U(1)_{B-L}$ & $-1$  &  $-1$  &  $-1$
 & $\frac{1}{3}$&  $\frac{1}{3}$&  $\frac{1}{3}$& 0& 0 & -2 & 2
   \\ \hline
\end{tabular}
\caption{\label{ub-l-charge} The $U(1)_{B-L}$ charges of the superfields.}
\end{center}
\end{table}
\end{widetext}

For universal SUSY soft breaking terms at grand unification scale,
$M_X$, the soft breaking Lagrangian is given by %
\begin{widetext}\bea %
-{\cal L}_{soft} &=& m^2_0\left[|\widetilde Q_i|^2+|\widetilde
U_i|^2+|\widetilde D_i|^2+|\widetilde L_i|^2+|\widetilde E_i|^2
\right.
+ \left.|\widetilde N_i|^2 +|H_u|^2+| H_d|^2+|
\chi_1|^2+|\chi_2|^2\right] \nn
\\
&+& A_0\left[Y_{U}{\widetilde Q}{\widetilde
U}^cH_u+Y_{D}{\widetilde Q}{\widetilde D}^cH_d+Y_{E}{\widetilde
L}{\widetilde E}^cH_d + Y_{\nu}{\widetilde L}{\widetilde
N}^cH_u+Y_{N}{\widetilde N}^c{\widetilde N}^c\chi_1 \right] \\&+&
\left[B(\mu H_uH_d+\mu'\chi_1\chi_2)+h.c.\right]
+
\frac{1}{2}M_{1/2}\left[{\widetilde g}^a{\widetilde
g}^a+{\widetilde W}^a{\widetilde W}^a+{\widetilde B}{\widetilde
B}+{\widetilde Z}_{B-L}{\widetilde Z}_{B-L}+h.c.\right],\nn
\eea %
\end{widetext}
where the tilde denotes the scalar components of the chiral matter
superfields and fermionic components of the vector superfields.
The scalar components of the Higgs superfields $H_{u,d}$ and
$\chi_{1,2}$ are denoted as  $H_{u,d}$ and $\chi_{1,2}$,
respectively.

As shown in Ref. \cite{s.khalil-a.masiero}, both $B-L$ and
electroweak (EW) symmetries can be broken radiatively in the
supersymmetric theories. 
In this class of models, the EW, $B-L$ and soft SUSY breaking are
related and occur at the TeV scale. 
The conditions for the EW symmetry breaking are given by %
\bea
\mu^2&=&\frac{m^2_{H_d}-m^2_{H_u}\tan^2\beta}{\tan^2\beta-1}-M^2_Z/2,~
\sin2\beta=\frac{-2m^2_3}{m^2_1+m^2_2}\label{sm-higgs-condition},\nn\\
\eea
 where
 \bea
&& m^2_i=m^2_0+\mu^2,~~i=1,~2~~~~~m^2_3=-B\mu,~~~~~\tan\beta=\frac{v_u}{v_d},
\nn\\&&
<H_u>=v_u/\sqrt2,~~~<H_d>=v_d/\sqrt2.
 \eea
Here $m_{H_u}$ and $m_{H_d}$ are the SM-like Higgs masses at the
EW scale. $M_Z$ is a neutral gauge boson in the SM. It is worth
noting that the breaking $SU(2)_L\times U(1)_Y$ occurs at the
correct scale of the charged gauge boson ($M_W\sim80$ GeV).
Similarly, the conditions for the $B-L$ radiative symmetry
breaking are given by \cite{s.khalil-q.shafi}
\bea
\mu^{'2}&=&\frac{\mu^2_{1}-\mu^2_{2}\tan^2\theta}{\tan^2\theta-1}-M^2_{Z_{B-L}}/2,~
\sin2\theta=\frac{-2\mu^2_3}{\mu^2_1+\mu^2_2},\label{b-l-higgs-condition}\nn\\
\eea
 where
 \bea
&& \mu^2_i=m^2_0+\mu^{'2},~~i=1,~2~~~~~\mu^2_3=-B\mu',~~~~~\tan\theta=\frac{v'_1}{v'_2},\nn\\&&
<\chi_1>=v'_1/\sqrt2,~~~<\chi_2>=v'_2/\sqrt2.
\eea
  Here $m_{\chi_1}$ and $m_{\chi_2}$ are the $U(1)_{B-L}$-like Higgs masses at the TeV scale.
The key point for implementing the radiative $B-L$ symmetry
breaking is that the scalar potential for $\chi_1$ and $\chi_2$
receives substantial radiative corrections. In particular, a
negative squared mass would trigger the $B-L$ symmetry breaking of
$U(1)_{B-L}$. It was shown that the masses of Higgs singlets
$\chi_1$ and $\chi_2$ run differently in the way that
$m^2_{\chi_1}$ can be negative whereas  $m^2_{\chi_2}$ remains
positive. The renormalization group equation (RGE) for the $B-L$
couplings and mass parameters can be derived from the general
results for SUSY RGEs of Ref. \cite{martin-vaughn}. After $B-L$
symmetry breaking, the $U(1)_{B-L}$ gauge boson acquires a
mass~\cite{Khalil:2006yi}: $M^2_{Z_{B-L}}=4g^2_{B-L}v'$. The high
energy experimental searches for an extra neutral gauge boson
impose lower bounds on this mass. The most stringent constraint on
$U(1)_{B-L}$ obtained from LEP ll result, which implies
\cite{m.carena} \be \frac{M_{Z_{B-L}}}{g_{B-L}}>6
~TeV.\label{z-b-l-constrain} \ee


Now we analyze mass-spectrums which have some deviations from
MSSM-spectrums in particular, SM singlet Higgs bosons, the
right-handed sneutrinos, and the neutralinos.  The Higgs sector in
the SUSY $B-L$ extension of the SM consists of two Higgs doublets
and two Higgs singlets with no mixing. However, after the $B-L$
symmetry breaking, one of the four degrees of freedom contained in
the two complex singlet $\chi_1$ and $\chi_2$ are swallowed by the
$Z^0_{B-L}$ to become massive. Therefore, in addition to the usual
five MSSM Higgs bosons: neutral pseudoscalar Higgs bosons $A$, two
neutral scalars $h$ and $H$ and a charged Higgs boson $H^{\pm}$,
three new physical degrees of freedom remain
\cite{s.khalil-a.masiero}. They form a neutral pseudoscalar Higgs
boson $A'$ and two neutral scalars $h'$ and $H'$. Their masses at
tree level
are given by %
\begin{widetext}
\begin{equation}
m_{A'}^2 = \mu_1^2 + \mu_2^2,
~~~
m_{H',h'}^2 = \frac{1}{2}\left(m_A'^2 + M^2_{Z_{B-L}} \pm
\sqrt{(m_{A'}^2 + M_{Z_{B-L}}^2)^2-4m_A^{'^2} M_{Z_{B-L}}^2\cos 2
\theta}\right).
\end{equation}\end{widetext}
The physical CP-even extra-Higgs bosons are obtained from the
rotation of angle $\alpha$: %
\be%
\left(\begin{array}{c}h'\\
H'\end{array} \right) = \left(\begin{array}{cc} \cos\alpha &
\sin\alpha \\
-\sin\alpha & \cos\alpha \end{array}\right) \left(\begin{array}{c}
\chi_1 \\
\chi_2 \end{array}\right), %
\label{mixing-matrix}
\ee %
where the mixing angle $\alpha$ is given by%
\be%
\alpha=\frac{1}{2} \tan^{-1}\left[\tan2\theta \frac{M_A'^2 +
M_Z'^2}{M_A'^2 - M_Z'^2} \right]. %
\label{mixing-angle}
\ee %
For $v'_1>>v'_2$, one finds the mixing angle $\alpha$ is very
small, hence the above diagonalizing matrix is close to the
identity. In this case, to a good approximation, one can assume
that $h'\equiv \chi_1$ and $H'\equiv \chi_2$. We are going to
adopt this assumption here.

Now we turn to the right-handed sneutrinos, in the basis of
$(\phi_{\nu_L},~\phi_{\nu_R})$ with
$\phi_{\nu_L}=(\widetilde{\nu}_L,~{\widetilde\nu}^{*}_L)$ and
$\phi_{\nu_R}=(\widetilde{\nu}_R,~{\widetilde\nu}^{*}_R)$, the
sneutrino mass matrix is given by the following $12\times12$
hermitian matrix: \be {\cal M}^2=\left(\begin{array}{cc}
M^2_{\nu_L\nu_L} &M^2_{\nu_L\nu_R}\\
 M^2_{\nu_R\nu_L} & M^2_{\nu_R\nu_R}\\
\end{array}\right),
\ee
where $M^2_{\nu_A\nu_B}(A,B=L,R)$ can be written as
 \be
M^2_{\nu_A\nu_B}
=
\left(\begin{array}{cc}
M^2_{A^{\dagger}B} & M^{2*}_{A^{T}B} \\
M^{2}_{A^{T}B} & M^{2*}_{A^{\dagger}B}\\
\end{array}\right),
\ee
with %
\begin{widetext}\bea %
M^2_{\nu^{\dagger}_L\nu_L} &=&
U^{\dagger}_{MNS}m^2_0U_{MNS}+\frac{M^2_Z}{2}\cos2\beta+
v^2\sin^2\beta U^{\dagger}_{MNS}(Y^{\dagger}_{\nu}Y_{\nu})U_{MNS},
\nn\\
M^2_{\nu^{\dagger}_R\nu_R}
&=&
m^2_0+M^2_N,
\nn\\
M^2_{\nu^{T}_L\nu_R}
&=&
v\sin\beta U^{\dagger}_{MNS}A_0(Y_N)^{\dagger}
+
v\cos\beta\mu U^{\dagger}_{MNS}A_0(Y_{\nu})^{\dagger} ,
\nn\\
M^2_{\nu^{T}_R\nu_R}
&=&
v'\sin\theta A_0(Y_N)^{\dagger} ,
\nn\\
M^2_{\nu^{\dagger}_L\nu_R}
&=&
v\sin\beta A_0(Y_{\nu})M_{N},
\nn\\
M^2_{\nu^{T}_L\nu_L} &=& 0, %
\eea %
\end{widetext}
where $v'\sin\theta=<\chi_1>$,
$M_N=Y_Nv'$ and $U_{MNS}$ is $3\times3$ unitary matrix termed the
Maki-Nakagawa-Sakata lepton mixing matrix \cite{mns}. Therefore, in
general the order of magnitude of the sneutrino mass matrix is as
follows: \be {\cal M}^2\sim \left(\begin{array}{cc}
{\cal O}(v^2) & {\cal O}(vv') \\
{\cal O}(vv') &{\cal O}(v^{'2})\\
\end{array}\right).
\ee Since $v'\sim$ TeV, the sneutrino matrix elements are of the
same order and there is no seesaw type behavior as usually found
in MSSM extended with heavy right-handed neutrinos. Therefore a
significant mixing among the left- and right- handed sneutrinos is
obtained. The phenomenological consequences for such mixing have
been studied in \cite{gross-haber}.

Finally, we consider the neutralino sector. The  neutral
gaugino-higgsino mass matrix can be written as: %
\bea
&&{\cal M}_7({\widetilde B},~{\widetilde W}^3,~{\widetilde
H}^0_d,~{\widetilde H}^0_u,~{\widetilde \chi_1},~{\widetilde
\chi_2},~{\widetilde Z}_{B-L}) \equiv \left(\begin{array}{cc}
{\cal M}_4 & {\cal O}\\
 {\cal O}^T &  {\cal M}_3\\
\end{array}\right),\nn\\
\eea%
where the ${\cal M}_4$ is the MSSM-type neutralino mass matrix and
${\cal M}_{3}$ is $3\times 3$ additional neutralino mass matrix,
which is given by:%
\begin{widetext}\be%
{\cal M}_3 = \left(\begin{array}{ccc}
0 &  -\mu' & -2g_{B-L}v'\sin\theta \\
-\mu' & 0 & 2g_{B-L}v'\cos\theta  \\
-2g_{B-L}v'\sin\theta & ~~~~~~~ 2g_{B-L}v'\cos\theta & M_{1/2}\\
\end{array}\right).
\label{mass-matrix.1} \ee\end{widetext}
As a feature of the orthogonality of $U(1)_Y$ and $U(1)_{B-L}$ in
this class of models, there is no mixing between ${\cal M}_4$ and
${\cal M}_3$ at tree level. Note that in extra $U(1)$ gauged
models, which are proposed to provide an explanation for the TeV
scale of $\mu$-term through the vev of a singlet scalar, the
neutralino mass matrix is given by $6 \times 6$ matrix. If the the
extra singlet fermion is the lightest neutralino, then it can be
an interesting candidate for dark matter, as shown in
Ref.\cite{jarecka-kalinowski}.
In our case, one diagonalizes the real matrix ${\cal M}_{7}$ with
a symmetric mixing matrix $V$ such as
\be V{\cal
M}_7V^{T}=diag.(m_{\chi^0_k}),~~k=1,..,7.\label{general} \ee In
this aspect, the lightest neutralino (LSP) has the following
decomposition \be \chi^0_1=V_{11}{\widetilde B}+V_{12}{\widetilde
W}^3+V_{13}{\widetilde H}^0_d+V_{14}{\widetilde
H}^0_u+V_{15}{\widetilde \chi_1}+V_{16}{\widetilde
\chi_2}+V_{17}{\widetilde Z}_{B-L}. \ee The LSP is called pure
$\widetilde Z_{B-L}$ if $V_{17}\sim1$ and $V_{1i}\sim0$,
$i=1,..,6$ and pure $\widetilde\chi_{1(2)}$ if $V_{15(6)}\sim1$
and all the other coefficients are close to zero. In our analysis,
we will focus on these two types of LSP and analyze their
potential contributions to DM in the universe.
The mass eigenstates of the matrix ${\cal M}_{3}$ are  in general
nontrivial mixtures of the fermions $(\widetilde
\chi_1,~\widetilde \chi_2,~\widetilde Z_{B-L}$). The limit of pure
$\widetilde Z_{B-L}$ that we consider can be obtained if
$v'<<\mu'$ and the limit of pure $\widetilde \chi_{1(2)}$ can be
obtained if
$\mu',~v'\sin(\cos)\theta<<v'\cos(\sin)\theta$\footnote {We would
like to thank the referee for drawing our attention to this
point.}.\vspace{1.0cm}
\section{LSP Annihilation Cross Section in $U(1)_{B-L}$ SUSY Model}
%
As advocated in the previous section, we focus on the cases where
LSP is pure $\tilde Z_{B-L}$ or $\tilde\chi_{1(2)}$. In this case,
the relevant Lagrangian is given by
\begin{widetext}
\bea &&\hspace{-0.75cm}- {\cal L}_{\widetilde {Z}_{B-L}}
 \simeq
 i\sqrt2g_{B-L}Y^f_{B-L}{\overline{\widetilde Z}}_{B-L}P_Rf\widetilde f_L
 +i\sqrt2g_{B-L}Y^f_{B-L}{\overline{\widetilde Z}}_{B-L}P_Lf\widetilde f_R
  +c.c.,
 \label{b-l-lag-z-ino}
\\
&&\hspace{-0.75cm}- {\cal L}_{\widetilde {\chi}_{1}}  \simeq
i\sqrt2g_{B-L}
Y^{\chi_{1}}_{B-L}{\overline{\widetilde\chi}_1}\slash\!\!\!
Z_{B-L}\gamma_5{\widetilde\chi}_1 \!+\! i\sqrt2g_{B-L}
Y^{\chi_{1}}_{B-L}{\overline{\widetilde\chi}_1}{\widetilde
Z_{B-L}}\chi_1 \!+\!(Y_N)_{ij}{\widetilde\chi}_1{
N^c}_i{\widetilde N^c}_j
 +c.c., ~~~~\label{b-l-lag-chi1-ino}
 \\
&& \hspace{-0.75cm} -{\cal L}_{\widetilde {\chi}_{2}}
 \simeq i\sqrt2g_{B-L}
Y^{\chi_{2}}_{B-L}{\overline{\widetilde\chi}_2}\slash\!\!\!
Z_{B-L}\gamma_5{\widetilde\chi}_2 + i\sqrt2g_{B-L}
Y^{\chi_{2}}_{B-L}{\overline{\widetilde\chi}_2}{\widetilde
Z_{B-L}}\chi_2
 +c.c.,
 \label{b-l-lag-chi2-ino}
\eea\end{widetext}
where $f$ refers to all the SM fermions, including the
right-handed neutrinos. $\widetilde f_L$ and $\widetilde f_R$ are
the left-handed and right-handed sfermions mass eignstates
respectively. $Y^f_{B-L}$ is the $B-L$ charge defined in Table
\ref{ub-l-charge}. We assume the first right-handed neutrino $N_1$
is of order ${\cal O}(100)$ GeV, therefore the annihilation
channel of the LSP into $N_1 N_1$ is also considered.

\begin{figure}[htb]
\unitlength=1mm \hspace{-7cm}
\begin{picture}(35,25)
\includegraphics[width=9cm]{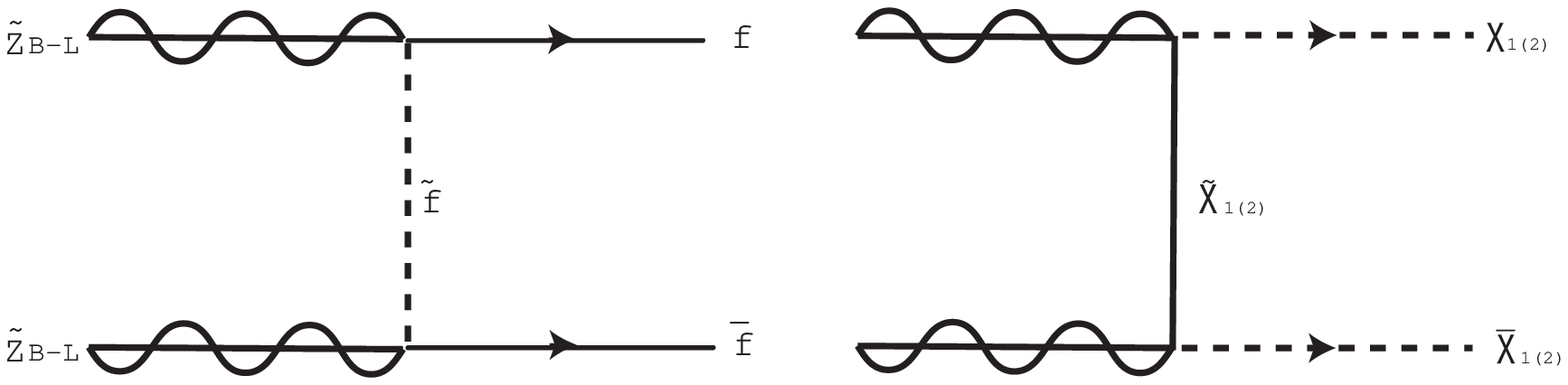}
\end{picture}
\caption{The dominant annihilation cross sections in the case of
the $\widetilde{Z}_{B-L}$-like LSP. Note that u-channel is also taken
into consideration for each of diagram.} \label{z-b-l-decay}
\end{figure}
From Eq. (\ref{b-l-lag-z-ino}), one finds that the dominant
annihilation processes of $\chi^0_1\equiv \widetilde Z_{B-L}$ are
given in Figure \ref{z-b-l-decay}. Our computation for the
annihilation cross section leads to the following $a_{\widetilde
Z_{B-L}}$ and $b_{\widetilde Z_{B-L}}$, where the approximation
$\langle \sigma_{ann} v \rangle \simeq a + b~ v^2$, with $v$ is
the
velocity of the incoming LSP, is assumed:%
\begin{widetext}\bea a_{\widetilde Z_{B-L}} &=& \frac{g^4_{B-L}}{54\pi
m^2_{\chi^0_1}} \L[\beta'_tr^2_tz^2_t+27\beta'_Nr^2_Nz^2_N\R],
\\
b_{\widetilde Z_{B-L}} &=& \frac{167g^4_{B-L}\beta'_fr^2_f}{162\pi
m^2_{\chi^0_1}}(1-2r_f+2r^2_f)
+
\frac{g^4_{B-L}}{4\pi m^2_{\chi^0_1}}
\L[\frac{\beta'_tr^2_t}{27}\{a_1+r_1+z^2(a_4+r_4)\}_t+\beta'_tr^2_t\{a_1+r_1+z^2(a_4+r_4)\}_N\R]
\nn\\
&+&
\frac{4g^4_{B-L}}{\pi m^2_{\chi^0_1}}|O_{in}O^T_{im}|^2|V_{2,i+4}V^T_{2,i+4}|^2
\beta'_{\chi}r^2_{\chi}\L[\frac{4}{3}(1+w^2_{\chi^0_2})\beta'^2_{\chi}-1\R]
, ~~~(i=1~or~2)
\label{chi-cross-section}
\\
a_1&=&\frac{2}{3}+\frac{1}{4}x_a^2 z_a^2-\frac{5}{12}z_a^2 ,~
a_4=\frac{x_a^2-3}{4},~
r_1=\frac{r_a}{3}\L[-4+z_a^2+r_a(4-3z_a^2-z_a^4)\R],~r_4=\frac{r_a}{3}(-3+2z_a^2+5r_a\beta_a'^2),
\nn\\
z_a&=&m_a/m_{\chi^0},~~w_{\alpha}=m_{\alpha}/m_{\chi^0}, ~~
r_{a}=(1-z^2_a+w^2_{\alpha})^{-1},~~
\beta'^2_{a}=1-z^2_a,~~x^2_a=\frac{z^2_a}{2(1-z^2_a)}.
\label{thermal-chi}
\eea\end{widetext}%
where $m_a$ is a final-state mass, $m_{\alpha}$ is a
mediated-particle mass, $O$ is the extra Higgs mixing matrix, as
defined in Eq. (\ref{mixing-matrix}) and Eq. (\ref{mixing-angle}).
In our approximation, $O_{in}$ is given by $O_{in}=\delta_{in}$,
and we set $m_{\tilde f}\equiv m_{\tilde f_R}\simeq m_{\tilde
f_L}$. Moreover, $V_{2i}$ is the coefficient of Next LSP (NLSP).
We assume that $\widetilde{\chi}_{1(2)}$ is our NLSP, therefore
$V_{2,i+4}\simeq1(i=1~or~ 2),~V_{2j}\simeq0(j\neq 5~{\rm or}~6)$.
In the range of parameter space that we consider, the values of
$a_{\widetilde Z_{B-L}}$ and $b_{\widetilde Z_{B-L}}$ are
typically $\lsim 10^{-8}$. For $m_{\widetilde Z_{B-L}} \gsim 100$
GeV, the annihilation channels into extra-Higgs $\chi_1$ and
$\chi_2$ may give the dominant contributions to $b_{\widetilde
Z_{B-L}}$.


\begin{figure}[htb]
\unitlength=1mm \hspace{-5cm}
\begin{picture}(30,20)
\includegraphics[width=10cm]{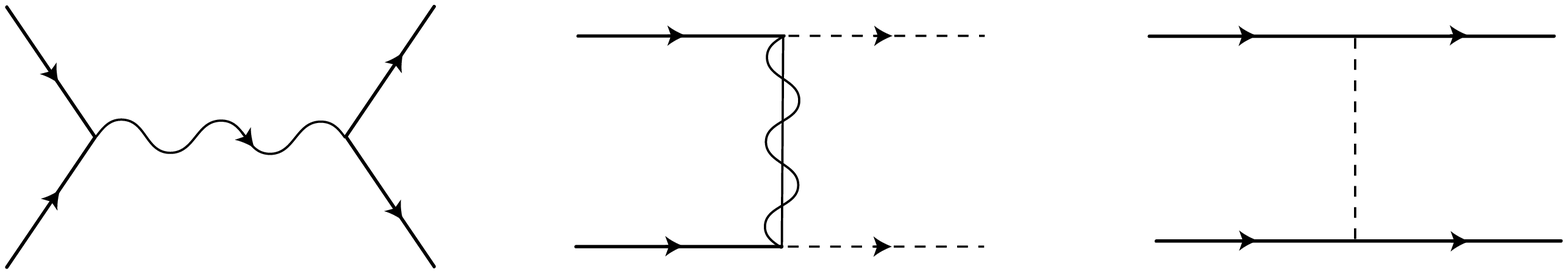}
\put(-105,17){$\tilde{\chi}_{1}$ }\put(-105,0){$\tilde{\chi}_{1}$ }\put(-90,4){$Z_{B-L}$ }\put(-70,17){$\bar{f}$ }\put(-70,0){$f$ }
%
\put(-47,7){$\tilde Z_{B-L}$ }\put(-36,15){$\chi_{1(2)}$ }\put(-36,0){$\bar{\chi}_{1(2)}$ }
%
\put(-12,7){$\widetilde N$ }\put(2,14){$N$ }\put(2,0){$\bar{N}$ } 
\end{picture}
\caption{The dominant annihilation channels of the
$\tilde{\chi}_1$-like LSP. For the last two diagrams, the
$u$-channel is also considered. } \label{chi-1-decay}
\end{figure}


Now we turn to the Higgsino contributions. From Eq.
(\ref{b-l-lag-chi1-ino}), one finds that the dominant annihilation
processes of $\chi^0_1\equiv \tilde \chi_{1}$ are given in Figure
\ref{chi-1-decay}. The computation of the cross section leads to
the following results for
$a_{\widetilde \chi_{1}}$ and $b_{\widetilde\chi_{1}}$:  %
\begin{widetext}\bea%
a_{\widetilde\chi_{1}} &=&
\frac{\beta'_Nz^2_N}{\pi}\L[\frac{(Y_{N,1m})^4r^2_N}{32m^2_{\chi^0_1}}
+
\frac{g^4_{B-L}m^2_{\chi^0_1}}{m^4_{Z_{B-L}}}\L(1-4\frac{m^2_{\chi^0_1}}{m^2_{Z_{B-L}}}\R)^{-2}
+
\frac{\sqrt2(Y_{N,1m})^2g^2_{B-L}}{4m^2_{Z_{B-L}}}\L(1-4\frac{m^2_{\chi^0_1}}{m^2_{Z_{B-L}}}\R)^{-1}\R],
\nn\\\\
b_{\widetilde\chi_{1}}
&=&
\frac{4g^4_{B-L}}{3\pi}\frac{m^2_{\chi^0_1}}{M^4_{Z_{B-L}}}
\L[\frac{23}{3}+\frac{1}{2}(1-z^2_t)\L(\frac{2}{3}+\frac{1}{4}x^2_tz^2_t-\frac{5}{12}z^2_t\R)
\R]
\L(1-4\frac{m^2_{\chi^0_1}}{M^2_{Z_{B-L}}}\R)^{-2}
\nn\\
&+&
\frac{\beta'_N}{\pi}\L[\frac{(Y_{N,1m})^4r^2_N}{32m^2_{\chi^0_1}}(a_1+r_1)_N
+
\frac{g^4_{B-L}m^2_{\chi^0_1}}{m^4_{Z_{B-L}}}\L(1-4\frac{m^2_{\chi^0_1}}{m^2_{Z_{B-L}}}\R)^{-2}a_{1N}\R]
\nn\\
&+& \frac{\beta'_N}{\pi}\L[
\frac{\sqrt2(Y_{N,1m})^2g^2_{B-L}}{4m^2_{Z_{B-L}}}\L(1-4\frac{m^2_{\chi^0_1}}{m^2_{Z_{B-L}}}\R)^{-1}(a_1+r_5z^2-\frac{2}{3}r\beta'^2)_N\R]
\nn\\
&+& \frac{4g^4_{B-L}}{\pi
m^2_{\chi^0_1}}|O_{1n}O^T_{1m}|^2|V_{27}V^T_{27}|^2
\beta'_{\chi}r^2_{\chi}\L[\frac{4}{3}(1+w^2_{\chi^0_2})\beta'^2_{\chi}-1\R].
\label{chi1-cross-section} %
\eea\end{widetext}%
Here $V_{27}$ is the coefficient of NLSP. We assume that
$\widetilde{Z}_{B-L}$ is our NLSP, therefore $V_{27}\simeq 1, ~
V_{2i}\simeq 0 (i\neq7)$. We also assume $(Y_N)_{1m}\simeq
(Y_N)_{11}$.

\begin{figure}[htb]
\unitlength=1mm \hspace{-7cm}
\begin{picture}(30,35)
\includegraphics[width=9cm]{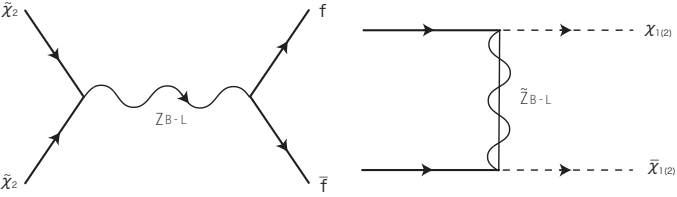}
%
\end{picture}
\caption{The dominant annihilation cross section in case of the
$\tilde{\chi}_2$-like LSP. Note that u-channel is also taken into
consideration for the t-channel diagram.} \label{chi-2-decay}
\end{figure}
Finally we consider the annihilation process of $\chi^0_1\equiv
\tilde \chi_{2}$. From Eq. (\ref{b-l-lag-chi2-ino}), one finds
that $\tilde \chi_{2} \tilde \chi_{2}$ annihilation is dominated
by the diagrams in Figure \ref{chi-2-decay}. The computation to
the cross section of $\widetilde \chi_{2}$ leads to $a_{\widetilde
\chi_{2}}=0$ and $b_{\tilde\chi_{2}}$ is given by%
\begin{widetext}\bea%
b_{\widetilde\chi_{2}} &=&
\frac{4g^4_{B-L}}{3\pi}\frac{m^2_{\chi^0_1}}{M^4_{Z_{B-L}}}
\L[\frac{23}{3}+\frac{1}{2}(1-z^2_t)\L(\frac{2}{3}+\frac{1}{4}x^2_tz^2_t-\frac{5}{12}z^2_t\R)
\R] \L(1-4\frac{m^2_{\chi^0_1}}{M^2_{Z_{B-L}}}\R)^{-2}
\nn\\&+&
 \frac{4g^4_{B-L}}{\pi
m^2_{\chi^0_1}}|O_{2n}O^T_{2m}|^2|V_{27}V^T_{27}|^2
\beta'_{\chi}r^2_{\chi}\L[\frac{4}{3}(1+w^2_{\chi^0_2})\beta'^2_{\chi}-1\R].
\label{chi2-cross-section} %
\eea\end{widetext}

It is remarkable that for $m_{\widetilde\chi_{1,2}}\gsim 100$ GeV,
their annihilations are dominated by extra-Higgs channel.
Therefore, $b_{\widetilde\chi_{1}}$ is very close to
$b_{\widetilde\chi_{2}}$ and $a_{\widetilde\chi_{1}}$ is quite
suppressed. Thus, in this region of parameter space both
$\widetilde\chi_{1}$ and $\widetilde\chi_{2}$ have very similar
annihilation cross section values.

\section{Constraints from Muon Anomalous Magnetic Moment}
\begin{figure}[htb]
\unitlength=1mm
\hspace{-5.5cm}
\begin{picture}(25,35)
\includegraphics[width=7cm]{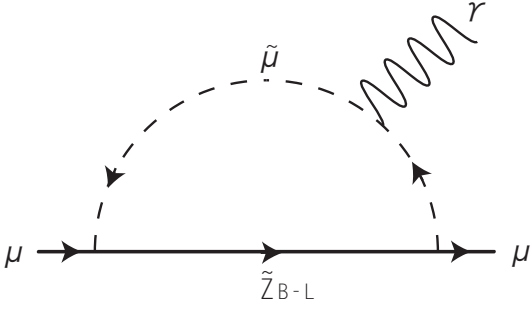}
\end{picture}
\caption{$\tilde Z_{B-L}$ contribution to the muon anomalous
magnetic dipole moment} \label{g-2}
\end{figure}

In the case of $\tilde Z_{B-L}$-like LSP, a significant
contribution to muon anomalous magnetic moment $(a_{\mu})$ may be
obtained due to the 1-loop diagram mediated by $\tilde Z_{B-L}$
and smuon, as shown in Figure \ref{g-2}. Note that
$\widetilde\chi_{1,2}$ have no direct couplings with the SM
fermions, thus they do not contribute to $a_{\mu}$.
The recent experimental value has been determined with a very high
precision by the E821 Collaboration at the National laoratory
\cite{bennett} \be a^{exp.}_{\mu}=(116592080\pm 60)\times
10^{-11}. \ee This value differs from the SM predicion by the
following: \be \Delta a_{\mu}=a^{exp.}_{\mu}-a^{SM}_{\mu}=(278\pm
82)\times 10^{-11}.
\label{exp-g-2} %
\ee%
Therefore, $\tilde Z_{B-L}$ contribution to $a_{\mu}$ should
satisfy the following constraints: \be 1.96\times 10^{-9}\le
\Delta a^{\tilde Z_{B-L}}_{\mu} \le3.6\times 10^{-9}.
\label{exp-g-2} \ee Our computation for $\tilde Z_{B-L}$
contribution to $a_{\mu}$ leads to the following result: %
\begin{widetext}\bea \Delta a^{\tilde Z_{B-L}}_{\mu} &=&
-\frac{g^2_{B-L}}{8\pi^2}\sum_{i=1,2}
\frac{m_{\mu}}{6m^2_{\tilde\mu_i}(1-s_i)^4}\times
[{\sqrt
s_i}(1-s_i)(U_{\tilde\mu})_{2i}(U_{\tilde\mu})_{1i}Y^l_{B-L}Y^{E}_{B-L}6(1-s^2_i+2s_i\ln
s_i) \nn\\&+&
\frac{m_{\mu}}{m_{\tilde\mu_i}}(|(U_{\tilde\mu})_{2i}Y^E_{B-L}|^2+|(U_{\tilde\mu})_{1i}Y^{l}_{B-L}|^2)(1-6s_i+3s^2_i+2s^3_i-6s^2_i\ln
s_i)],\label{g-2-eq} %
\eea \end{widetext}%
where $U_{\tilde\mu}$ is a diagonalized unitary matrix of the
slepton sector, $s_i=(m_{\chi^0}/m_{m_{\tilde\mu_i}})^2$ and
$Y^{l(E)}_{B-L}$ is the $U(1)_{B-L}$ charge in the Table
\ref{ub-l-charge}. This result is consistent with the derivation
of the new contribution to $a_{\mu}$ in supersymmetric $U(1)'$
model \cite{g-2}.

Here few comments are in order: (i) The second term in $\Delta
a^{\tilde Z_{B-L}}_{\mu}$ is suppressed by $m_{\mu}/m_{\tilde
\mu_i}\simeq {\cal O}(10^{-3})$, while the first term is
proportional to the off-diagonal elements of the diagonalized
matrix $U_{\tilde\mu}$ which are typically of order ${\cal
O}(10^{-2})$. Therefore the first term is Eq. (\ref{g-2-eq}) gives
the dominant contribution to $\Delta a_{\mu}$. (ii) From the Eq.
(\ref{exp-g-2}), the sign of $\tilde Z_{B-L}$ contribution to
$\Delta a_{\mu}$ should be positive. Thus
$[(U_{\tilde\mu})_{11}(U_{\tilde\mu})_{21}+(U_{\tilde\mu})_{12}(U_{\tilde\mu})_{22}]Y^l_{B-L}Y^{E}_{B-L}$
must be negative. Note that $s_i<1$, hence the function
$f(s_i)={\sqrt s_i}(1-s_i)(1-s^2_i+2s_i\ln s_i)$ is always
positive. The elements of $U_{\tilde\mu}$ have a sign difference
that helps in satisfying this requirement and allows for positive
contribution to $\Delta a_{\mu}$. For example, in case
$m_{\tilde\mu L}= m_{\tilde\mu R}=A\simeq300$ GeV, $\mu= 500$ GeV
and $\tan\beta=10$, the corresponding $U_{\tilde\mu}$ matrix is
given by %
\bea
U_{\tilde\mu} &\sim& \left(\begin{array}{cc}
\bf{-1/\sqrt2} & \bf{1/\sqrt2}\\
\bf{1/\sqrt2} &\bf{1/\sqrt2}\\
\end{array}\right).
\eea %
(iii) Large values of $\tan\beta$ enhance the off-diagonal
elements of $U_{\tilde\mu}$. Hence $\Delta a^{\tilde
Z_{B-L}}_{\mu}$ are enhanced by large values of $\tan\beta$.

\begin{figure}
\unitlength=1mm
\hspace{-5.5cm}
\begin{picture}(25,50)
\includegraphics[width=8cm]{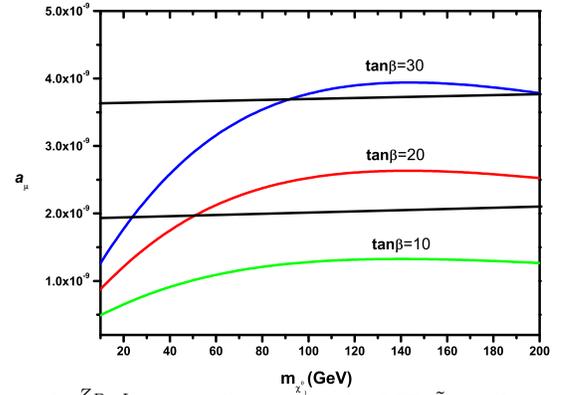}
\end{picture}
\vspace{-0.9cm}
 \caption{$\Delta a^{Z_{B-L}}_{\mu}$ versus the mass
of the LSP $\tilde Z_{B-L}$ for $m_{\tilde\mu_R}\simeq
m_{\tilde\mu_L}\simeq A\simeq300$ GeV, $\mu\simeq 500$ GeV with
$\tan\beta=10,~20$ ~and~$30$, and $g_{B-L}=0.5$.} \label{g-2-b-l}
\end{figure}

In Figure \ref{g-2-b-l}, we plot $\Delta a^{\tilde Z_{B-L}}_{\mu}$
as a function of the $\tilde Z_{B-L}$-like LSP mass,
$m_{\chi^0_1}$, for $\tan\beta=10,~20$ and $30$. Other SUSY
parameters are fixed as above. From this figure, it can be easily
seen that a significant $B-L$ contribution to $\Delta a_{\mu}$ can
be obtained for $\tan\beta>10$. For $\tan\beta=30$, the LSP mass
is constrained within the region $30$ GeV $< m_{\chi^0_1} < 100$
GeV. While for $\tan\beta=20$, the allowed region of
$m_{\chi^0_1}$ is rather wider: $m_{\chi^0_1} \gsim 60$ GeV.

\section{LSP Relic Abundance in $U(1)_{B-L}$ SUSY Model}

In this section, we compute the LSP relic abundance in
$U(1)_{B-L}$ SUSY Model. We adopt the standard computation of the
cosmological abundance, where the LSP is assumed to be in thermal
equilibrium with the SM particles in the early universe and
decoupled when it was non-relativistic. Therefore, the LSP density
can be obtained by solving the Boltzmann equation \cite{griest1}:
\be
\frac{dn_{\chi^0_1}}{dt}+3Hn_{\chi^0_1}=-<\sigma^{ann}_{\chi^0_1}v>
[(n_{\chi^0_1})^2-(n^{eq.}_{\chi^0_1})^2],\label{boltzmann} \ee
where $n_{\chi^0_1}$ is LSP number sensity with
$m_{\chi^0_1}=\rho_{\chi^0_1}n_{\chi^0_1}$. One defines $\Omega
_{\chi^0_1}=\rho_{\chi^0_1}/\rho_{c}$, where $\rho_{c}$ is the
critical mass density.
It turns out that \cite{griest1} \bea&&
 \Omega_{\chi^0_1}
h^2 \simeq \frac{8.76\times 10^{-11}GeV^{-2}}{g^{1/2}_{*
(T_F)}(a/x_F+3b/x^2_F)}, \label{relic-abundance} \nn\\&&
x_F=ln\frac{0.0955m_{pl}m_{{\chi^0}_1}(a+6b/x_F)}{(g_{*
(T_F)} x_F)^{1/2}}, \eea
where $m_{pl}$ is the Planck mass ($1.22\times10^{19}$ GeV) and
$g_{* (T_F)}$ enumerates the degrees of freedom of relativistic
particles at $T_F$. From the expressions, one notes that the LSP
relic abundance depends only on the LSP mass and the annihilation
cross section coefficients $a$ and $b$.

In our numerical calculation for the LSP annihilation cross
section, we consider the following values of masses for the
particles contributing in the process (extra-light Higgses
($\chi^0_{1(2)}$), sfermions ($\tilde f$), the lightest
right-handed neutrino ($N_1$) and the NLSP ($\chi^0_2$)): $
m_{\chi_{1(2)}}=100\ {\rm GeV}$, $\tilde f=200\ {\rm GeV}$,
$N_1=100\ {\rm GeV}$, $m_{\chi^0_2}=m_{\chi^0_1}+30\ {\rm GeV}$.
\begin{figure}
\unitlength=1mm
\hspace{-5.5cm}
\begin{picture}(25,65)
\includegraphics[width=8cm]{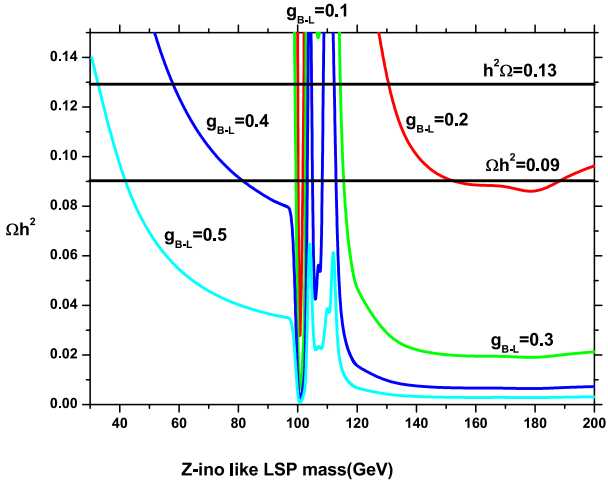}
\end{picture}
\vspace{-0.3cm} \caption{$\Omega h^2$ versus  $\widetilde
Z_{B-L}$-like LSP mass for $g_{B-L} \in [0.1, 0.5]$.}
\label{zb-l-relic}
\end{figure}


In Figure \ref{zb-l-relic}, we present the values of relic density
$\Omega h^2$ as a function of the LSP mass for $\widetilde
Z_{B-L}$--like LSP and $g_{B-L} \in[0.1,0.5]$. The horizontal
lines are experimentally allowed regions from the Wilkinson
Microwave Anisotropy Probe (WMAP) \cite{wmap} results for cold
dark matter relic density. Here, we have imposed the constraint on
the mass of $\widetilde Z_{B-L}$-like LSP: $m_{\widetilde Z_{B-L}}
\gsim 30$ GeV, due to the experimental limits on the muon
anomalous magnetic moment. From this figure, one notes that since
the sfermion mass is fixed at $200$ GeV, the annhiliation channel
due to its exchange produces a resonance at the LSP mass of order
$100\ {\rm GeV}$. For $g_{B-L}= 0.2$, the allowed region is rather
wide: $130\ {\rm GeV}< m_{\chi^0_1}$, while for $g_{B-L}= 0.3$,
the allowed region is reduced to around $120$ GeV. Finally, for
$g_{B-L} \gsim 0.4$ a lighter LSP ($m_{\tilde{Z}_{B-L}}\lsim 100$
GeV) is favored.

\begin{widetext}[htb]
\begin{figure}
\unitlength=1mm
\hspace{-15.5cm}
\begin{picture}(25,58)
\includegraphics[width=7.9cm]{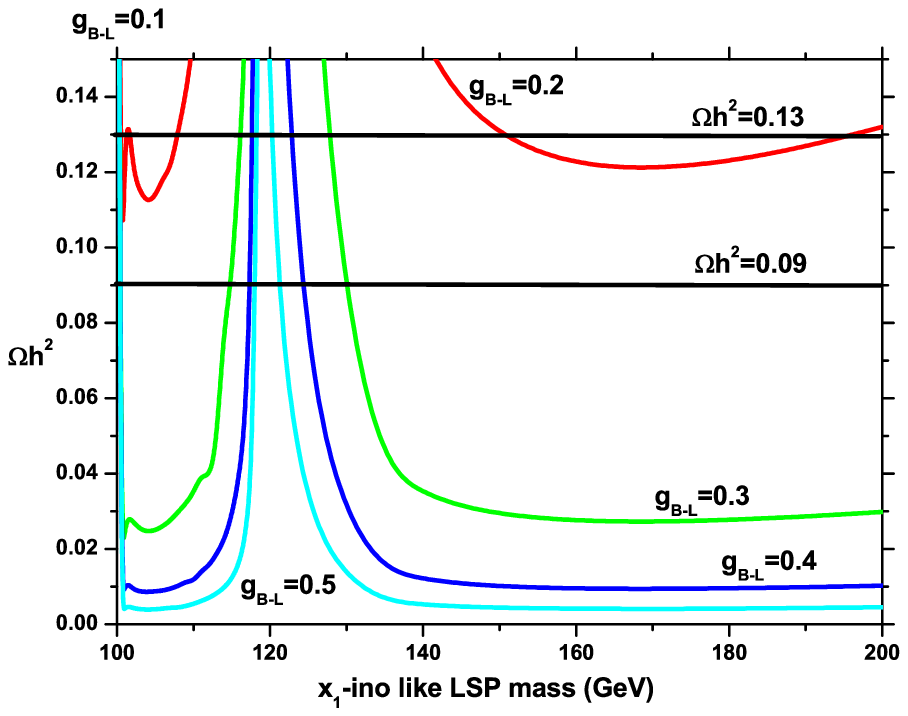}
\includegraphics[width=8cm]{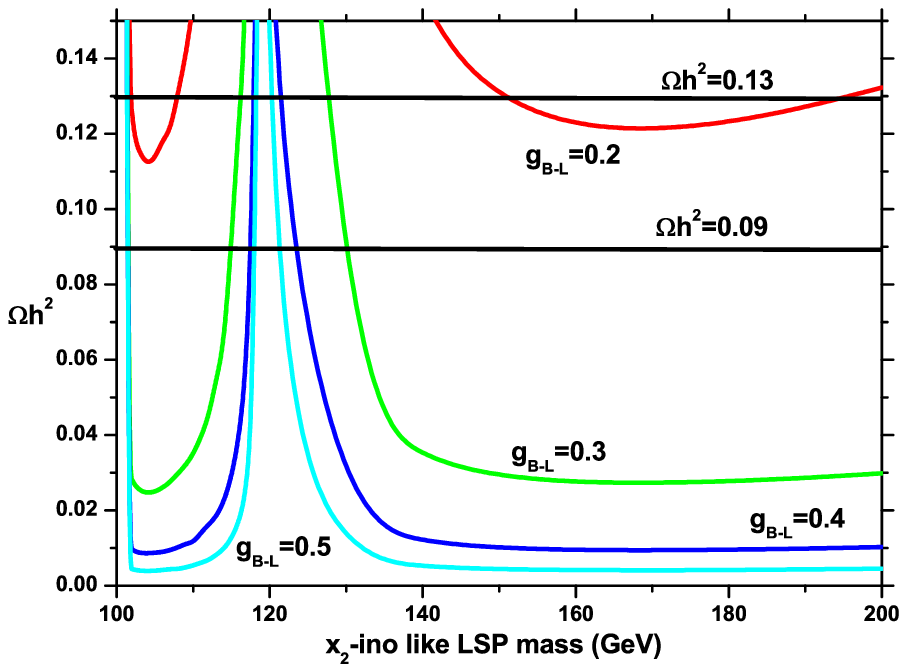}
\end{picture}
\vspace{-0.3cm} \caption{$\Omega h^2$ versus $\widetilde
\chi_{1,2}$-like LSP mass for $g_{B-L}\in [0.1, 0.5]$.}
\label{chi1-relic}
\end{figure}\end{widetext}

Now we turn to the Higgsino ${\widetilde \chi}_{1,2}$ LSP. In
Figure \ref{chi1-relic}, we plot the LSP relic density $\Omega
h^2$ as a function of ${\widetilde \chi}_{1}$ or ${\widetilde
\chi}_{2}$ mass.  As expected the relic abundance of ${\widetilde
\chi}_{1}$ or ${\widetilde \chi}_{2}$ are quite similar since they
have very close annihilation cross section. From this figure, one
notes that for $g_{B-L}\le 0.1$, there is no essentially any
allowed region due to the fact that the relic abundance becomes
quite large. While for $g_{B-L}= 0.2,0.3,0.4,0.5$, the allowed
regions for $m_{\chi_{1,2}}$  are given by $[150 - 190]$ GeV,
$[130 - 135]$ GeV, $[125 - 130]$ GeV, and $[115 - 120]$ GeV,
respectively.

\section{LSP Detection Rate in $U(1)_{B-L}$ SUSY Model}

In this section we analyze the effect of the event rates of our
relic neutralinos ($\widetilde Z_{B-L}, \widetilde \chi_{1(2)}$)
scattering off nuclei in terrestrial detectors. The direct
detection experiments provide the most natural way of searching
for the neutralino dark matters. The differential cross section
rate is given by \cite{skhalil-detec.} \be
\frac{dR}{dQ}=\frac{\sigma\rho_{\chi}}{2m_{\chi^0_1}m^2_r}|F(Q)|^2\int^{\infty}_{v_{min.}}\frac{f_1(v)}{v}dv,
\ee where$f_1(v)$ is the distribution of speeds relative to the
detector. The reduced mass is
$m_r=\frac{m_{\chi^0_1}m_N}{m_{\chi^0_1}+m_N}$, where $m_N$ is the
mass of the nucleus, $v_{min.}=\L(\frac{Qm_N}{2m^2_r}\R)^{1/2}$,
$Q$ is the energy deposited in the detector, and $\rho_{\chi^0_1}$
is the density of the neutralino near the Earth. It is common to
fix $\rho_{\chi^0_1}$ to be $\rho_{\chi^0_1}=0.3GeV/cm^3$. The
quantity $\sigma$ is the elastic-scattering cross section of the
LSP with a given nucleus. In our model, $\sigma$ has two
contributions: spin-independent (scalar) contribution due to the
squark exchange diagrams for ${\widetilde Z}_{B-L}$-like LSP, and
spin-dependent contribution arising from $Z_{B-L}$ gauge boson
exchange diagrams for ${\widetilde\chi}_{1(2)}$-like LSP. For
$^{76}Ge$ detector, where the total spin of $^{76}Ge$ is equal to
zero, we have a contribution from the scalar part only, which is
given by \be \sigma^{SI}=\frac{4m^2_r}{\pi}|Zf_p+(A-Z)f_n|^2,
\label{SI-cross-section} \ee where $Z$ is the nuclear charge, and
$A-Z$ is the number of neutrons. The expressions for the effective
couplings to proton and neutron, $f_p$ and $f_n$, can be found in
Ref. \cite{griest2}. Finally, the form factor $F(Q)$, in this case
is given by \cite{skhalil-detec.} \be
F^{SI}(Q)=\frac{3j_1(qR_1)}{qR_1}e^{\frac{-1}{2}q^2s^2}, \ee where
$q=\sqrt{2m_NQ}$ is the momentum transferred and $R_1$ is given by
$R_1=(R^2-5s^2)^{1/2}$ with $R=1.2fmA^{1/2}$ and $A$ is the mass
number of $^{76}Ge$. $j_1$ is the spherical Bessel function and
$s\simeq 1fm$.


For $^{73}Ge$ detector, where the total spin of $^{73}Ge$ is equal
to $J=\frac{9}{2}$, we have a contribution from spin-dependent
part only, which can be written as
\bea
&&\sigma^{SD}|F^{SD}(Q)|^2=\nn\\
&&\frac{4m^2_r}{2J+1}[(f^a_p)^2S_{pp}(q)+(f^a_n)^2S_{nn}(q)+f^a_pf^a_nS_{pn}(q)],
\label{SD-cross-section} \nn\\\eea
where
$S_{pp}(q)=S_{00}(q)+S_{11}(q)+S_{01}(q),~S_{nn}(q)=S_{00}(q)+S_{11}(q)-S_{01}(q)$
and $S_{pn}(q)=2[S_{00}(q)-S_{11}]$, and the expressions for
$f^a_p$ and $f^a_n$ can be found in Ref. \cite{gondolo}. The
values of the spin structure functions $S_{00}(q),~S_{11}(q)$ and
$S_{01}(q)$ are given in \cite{griest2}.

In case of ${\widetilde Z_{B-L}}$-like LSP, the effective
couplings to proton and neutron are very similar {\it i.e.}
$f_p\simeq f_n$. Therefore, the cross section,
$\sigma^{SI}\equiv\sigma^{SI}_{{\widetilde Z_{B-L}}}$, is given by
\be \sigma^{SI}_{{\widetilde Z_{B-L}}} \simeq
\frac{4m^2_r}{\pi}\L|\sum_{q} \frac{1}{2} <N|{\bar q}q|N>
\sum^{6}_{k=1}\frac{g_{{\tilde q}_{L_k} \chi q} g_{{\tilde
q}_{R_k} \chi q}}{m^2_{\tilde q_k}}\R|^2, \ee where $q$ refers to
$u,d,s,c,b,t$. The hadronic matrix elements are given by $<N|{\bar
q}q|N>=f^{p}_{T_q}m_p/m_q$. The values of the parameters
$f^{p}_{T_q}$ can be found in Ref. \cite{gondolo}. From Eq.
(\ref{b-l-lag-z-ino}), one finds that ${\widetilde Z_{B-L}}$
couples universary to all type of quarks, {\it i.e}. $g_{{\tilde
q}_{L_k} \chi q}= g_{{\tilde q}_{R_k} \chi q}\simeq i\sqrt 2
g_{B-L}Y^{q}_{B-L}$.
\begin{figure}
\unitlength=1mm
\hspace{-5.5cm}
\begin{picture}(25,55)
\includegraphics[width=7.9cm]{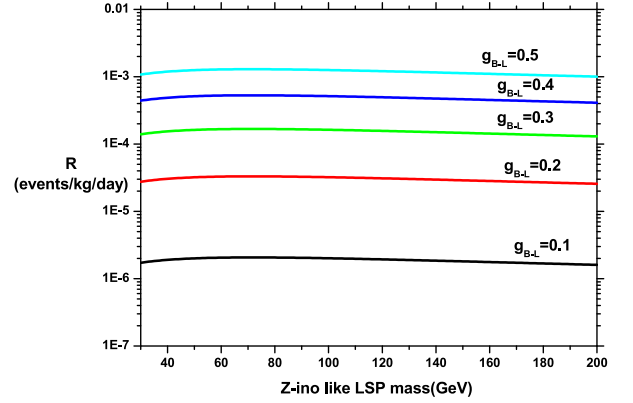}
\end{picture}
\vspace{-0.3cm} \caption{Detection rate versus  $\widetilde
Z_{B-L}$-like LSP mass for $g_{B-L} \in [0.1, 0.5]$.
As in previous figures $m_{\tilde q}=200$ GeV is assumed. }
\label{detec.zb-l}
\end{figure}

In Figure \ref{detec.zb-l}, we present our numerical results for the
event rate $R$ as a function of ${\widetilde Z_{B-L}}$-like LSP
mass for $m_{\tilde q}=200$ GeV and $g_{B-L}\in [0.1,0.5]$. As can
been seen from this figure, the detection rates are quite
sensitive to the value of gauge coupling $g_{B-L}$. This is due to
the fact that $R$ depends on the forth power of $g_{B-L}$.
Nevertheless, the detection rates are less than $10^{-3}$
events/kg/day, which are below the current experimental limit:
$0.01$ events/kg/day \cite{Ahmed:2008eu} . Thus, one can conclude
that $\widetilde{Z}_{B-L}$ is beyond the reach of near future
experiments.

Now we turn to the case of ${\widetilde \chi_{1(2)}}$-like LSP. As
mentioned above, in this case the scattering cross section is
given by the spin-dependent part:
$\sigma^{SD}\equiv\sigma^{SD}_{{\widetilde \chi_{1(2)}}}$, which
is given by Eq. (\ref{SD-cross-section}) with
\begin{widetext}\bea
f^a_N =
\sum_{q=u,d,s} (\Delta
q)_N\L(\frac{2g^2_{B-L}Y^{\chi}_{B-L}Y^{q}_{B-L}}{m^2_{Z_{B-L}}}\R)
\lsim \frac{2}{3}\L(\frac{1}{6000GeV}\R)^2\sum_{q=u,d,s} (\Delta
q)_N.
\eea\end{widetext}
 Here we have used the lower limit on the ratio:
$M_{Z_{B-L}}/g_{B-L}$ reported in Eq. (\ref{z-b-l-constrain}). The
numerical values of $S_{00}(q),~S_{11}(q),~S_{01}(q)$ and $(\Delta
q)_N$ can be found in Ref. \cite{gondolo}. From this expression,
it is clear that the detection rates of the extra Higgsinos-like
LSP are extremely small. They are typically less than $10^{-16}$
(events/kg/day). This result is consistent with the spin-dependent
contribution for the singlino in SUSY models with $U(1)'$
\cite{deCarlos:1997yv,langacker}. However, in this class of model,
unlike our $U(1)_{B-L}$ model, the singlino dominated LSP may
imply large detection rates, due to the spin independent
contributions.

\section{Conclusions}
We have studied the DM problem in supersymmetric $B-L$ extension
of the SM. We showed that the extra $B-L$ neutralinos (three extra
neutral fermions: $U(1)_{B-L}$ gaugino $\widetilde {Z}_{B-L}$ and
two Higgsinos $\widetilde{\chi}_{1,2}$) are interesting candidates
for cold DM. We provided analytic expressions for their
annihilation cross sections. We also computed the
$\widetilde{Z}_{B-L}$ contribution to muon anomolous magnetic
moment and showed that the current experimental limits impose a
lower bound of order $30$ GeV on $\widetilde{Z}_{B-L}$ mass. We
analyzed the thermal relic abundance of both $\widetilde
{Z}_{B-L}$ and $\widetilde{\chi}_{1,2}$. We showed that unlike the
LSP in MSSM, these particles can account for the measured relic
abundance with no conflict with other phenomenological
constraints. Finally, we discussed their direct detection rates
and showed that they are beyond the reach of our near future
experiments.

\section*{Acknowledgments} We thank D. A. Demir for
discussions. H.O. would like to thank Y. Daikoku, S. Nakamura and
Y. Kajiyama for interesting comments. This work was partially
supported by the ICTP grant Proj-30 and the Egyptian Academy for
Scientific Research and Technology.

\end{document}